\documentstyle[prb,aps,twocolumn,floats]{revtex}
\begin{document}
\draft
\twocolumn[\hsize\textwidth\columnwidth\hsize\csname@twocolumnfalse\endcsname
\title{Dynamics and Effective Actions of BCS Superconductors}
\author{Anne van Otterlo$^{1,2}$, Dmitrii S. Golubev$^{3}$,
	Andrei D. Zaikin$^{4,3}$, and Gianni Blatter$^{1}$}
\address{$^{1}$ Theoretische Physik, ETH-H\"{o}nggerberg,
	CH-8093 Z\"{u}rich, Switzerland\\
	$^{2}$ Physics Department, University of California,
	Davis, CA 95616, USA\\
	$^{3}$ I.E.Tamm Department of Theoretical Physics, P.N.Lebedev
	Physics Institute, Leninskii pr. 53, 117924 Moscow, Russia\\
        $^{4}$ Institut f\"{u}r Theoretische Festk\"orperphysik,
	Universit\"at Karlsruhe, 76128 Karlsruhe, FRG}
\maketitle
\begin{abstract}
We derive effective dynamical theories for metals and BCS superconductors,
based on the effective action formalism. Both the metallic regime
$T\gtrsim T_{C}$ and the superconducting regime $T\ll T_{C}$ are studied
in the clean and dirty limit. Furthermore, we consider the effect of
particle-hole asymmetry in the band structure. Using gauge invariance, the
electrodynamics of the problem is formulated in a transparent way. The
effective actions are useful starting points for treating dynamical problems
involving BCS superconductors.
\end{abstract}
\pacs{PACS numbers: 74.20.Fg, 71.10.-w}

]

\section{Introduction}
Dynamical problems in BCS theory~\cite{bcs} are diverse. They include the
electromagnetic response of superconductors~\cite{respo}, relaxation
phenomena, and collective modes in superconductors~\cite{gsch,ps}, e.g., the
Carlson-Goldman~\cite{carlg} and the Mooij-Sch\"{o}n~\cite{nice} mode.
Further examples are the motion of topological defects, e.g., vortex motion
in bulk samples~\cite{vorol,vorne,avo,volo}, quantum tunneling of
vortices~\cite{tunne}, and thermally activated or quantum phase slips in one
dimensional wires~\cite{lamh,zgoz}, as well as fluctuation effects, e.g.,
corrections to the conductivity above $T_{C}$~\cite{al,thomp,maki}, the
renormalization of the critical temperature and of the energy gap of low
dimensional dirty superconductors~\cite{fluct}, and the quantum melting
of the vortex lattice~\cite{qmelt}.
For all these phenomena, an effective (simple) theory of weak coupling BCS
superconductivity is desirable. However, such an effective theory is well
established only close to and above the critical temperature, where Time
Dependent Ginzburg-Landau (TDGL) theory -- although only under severe
additional restrictions -- can be derived from a microscopic starting
point~\cite{schmi}. Since the works of Abrahams and Tsuneto~\cite{abrts},
Popov~\cite{popov}, Kleinert~\cite{klein}, Ambegaokar~\cite{ambeg}, and
Stoof~\cite{stoof} some is known about the extensions to lower temperature,
but controversies concerning the subject persist~\cite{greit,schak,aitch,stone}.

Previous papers were mostly restricted to the clean limit at zero temperature
and neglected the Coulomb interaction~\cite{greit,schak,aitch,stone}. In the
present paper, we extend the existing literature in four ways. 1) We fully
account for the Coulomb interaction between electrons and between electrons
and the ionic Jellium background. 2) We also consider the dirty limit, in
which electrons move diffusively rather than ballistically. 3) We allow for
particle-hole asymmetry, i.e., the dependence of the density of states on
energy, as quantified by its derivative at the Fermi surface
$N'_{0}=\partial_{\epsilon}N(\epsilon_{F})$. The relevance of particle-hole
asymmetry for vortex motion and the corresponding Hall-effect has recently
been pointed out in Refs.~\cite{dors,avo}. 4) Finally, a guiding principle
is called for, as the expansion of the effective action that we will perform
(see below) is quite involved. We will make extensive use of gauge invariance
and the corresponding Ward-identities. Although perturbation expansions up to
a definite order may break gauge-invariance, the Ward identities allow us to
construct a manifestly gauge-invariant effective action within perturbation
theory.

In Refs.~\cite{greit,schak,aitch,stone} Galilei invariance in the clean limit
was used as a guiding principle. We argue that in real superconductors 
Galilei invariance can be broken by at least four mechanisms. 1) Scattering
on impurities and phonons break Galilei invariance in a trivial way.
2) Galilei invariance would require a perfect quadratic dispersion, which
is not realized in the bandstructure of usual crystal backgrounds.
3) The coupling to electromagnetism, which is Lorentz invariant, also
(weakly) breaks Galilei invariance (a $v_{F}/c$ effect). 4) At
nonzero temperature the gas of normal quasi-particle excitations provides
a preferred frame of reference which breaks Galilei invariance. 

Instead of Galilei invariance, we stress the role of gauge invariance.
The corresponding Ward identities express particle number and charge
conservation and can be used to rewrite the effective action in a
manifestly gauge-invariant way. This means that in the superconducting
phase the effective action depends only on the amplitude of the energy
gap $\vert\Delta\vert$, the superfluid velocity ${\bf v}_{s}=\frac{1}{2m}
({\bf \nabla}\varphi-\frac{2e}{c}{\bf A})$ and the chemical potential for
Cooper pairs $\Phi=V-\dot{\varphi}/2e$. This is in accord with the
Anderson-Higgs mechanism~\cite{ah}: the superconducting phase does not
appear explicitly in the effective action. We will see below that, for
example, if the equilibrium value of the energy gap $|\Delta_{0}|$ does
not depend on space and time coordinates and all fields vary slowly in
space and time, the action in the particle-hole symmetric case has the form
\begin{equation}
	S=\!\!\int\!\!\frac{{\,\mathchar'26\mkern-9mu{\!d}} q}{2}
	{\Big (}\frac{\epsilon{\bf E}^{2}
	+{\bf B}^{2}\!/\mu}{4\pi}\!+\!\chi_{A}\vert\Delta_{1}\vert^{2}
	\!+\!\chi_{L}{\bf v}^{2}_{s}\!+\!\chi_{J}\Phi^{2}{\Big )}\;.
\label{sintro}
\end{equation}
Here $|\Delta_{1}|$ is the small deviation of the order parameter from the
equilibrium value, the $\chi$'s are generalized susceptibilities, $\epsilon$
is the dielectric function of the metal, and $\mu$ is the magnetic
permeability. The terms involving ${\bf E}$ and ${\bf B}$ include the
electronic polarization contributions to the electromagnetic fields that
will be discussed below in Section III. The term involving the amplitude
fluctuations $\vert\Delta_{1}\vert$ is discussed in Section IV. Most
interesting are the last two terms: the first describes how the gradient
of the superconducting phase ${\bf \nabla}\varphi$ tries to adjust to the
local vector potential ${\bf A}$. The prefactor $\chi_{L}$ is proportional to
the superfluid density $n_{s}$ and this term is just the kinetic energy of
the superfluid, related to the DC Josephson effect and to London theory.
Similarly, the less familiar second term describes how the time derivative
of the superconducting phase $\dot{\varphi}$ tries to adjust to the local
scalar potential $V$. This is the term that produces the AC Josephson effect.
The prefactor $\chi_{J}$ is proportional to the superfluid density as well.
If the critical temperature $T_{C}$ is approached from below, the
superfluid density vanishes, and the chemical potential for Cooper pairs
$\Phi$ and the chemical potential for quasi-particles $V$ decouple.
This last term also describes how the motion of vortices in the mixed
state leads to a voltage drop across the sample via the time dependence
of the phase.

In case particle-hole symmetry is broken, an additional contribution to the
action arises, which is proportional to the derivative of the density of
states at the Fermi energy $N'_{0}=\partial_{\epsilon}N(\epsilon_{F})$.
The dimensionless parameter that characterizes the amount of particle-hole
symmetry breaking is $\gamma=\Delta N'_{0}/(2\lambda N^{2}_{0})$, with BCS
coupling constant $\lambda$. In usual weak coupling superconductors with
$\lambda N_{0}\sim 10^{-1}$ and $\Delta/\epsilon_{F}\sim 10^{-3}$ the
parameter $\gamma$ is rather small, $\gamma\sim 10^{-2}$. Nevertheless, it
is important for vortex motion, and its possible relation to the sign-anomaly
in the Hall effect has been discussed in Refs.~\cite{avo,dors}. The general
form of the particle-hole asymmetric part in the action is
\begin{equation}
	S=-2ieN_{0}\Gamma\int dx \; \Phi\left(\vert\Delta_{0}
	+\Delta_{1}\vert^{2}-\vert\Delta_{0}\vert^{2} \right)\;,
\label{spha}
\end{equation}
with $\Gamma=\gamma/\Delta_{0}$. The physical origin of this term is the
coupling of the electronic density to the energy gap when particle-hole
symmetry is broken. Thus, fluctuations in the amplitude of the gap cause
charge density fluctuations, which couple directly to the potential $\Phi$,
see also Section IV for a discussion of this point.

Our discussion below will be within the imaginary time Matsubara formalism. 
Here we will not address the important point of relaxation mechanisms like
spin-flip, electron-electron, and electron-phonon scattering that could be
accounted for. We will simply assume that some relaxation mechanism is
available which brings our superconducting system into equilibrium with a
big reservoir.
We would like to emphasize that this assumption is by no means in
contradiction with the main goal of our paper: to provide a convenient
approach for studying dynamical and nonequilibrium phenomena in
superconductors. Rather, it restricts the scope of the phenomena which can
be effectively described with our methods.

In what physical situations is our imaginary time formalism meant to work?
One such situation is quite standard: a superconductor only slightly driven
out of equilibrium, so that one can describe nonequilibrium effects within a
linear response theory and express the results in terms of equilibrium
correlation functions, which -- after a proper analytic continuation -- will
describe the dynamics of the system in real time. Another important class
of phenomena is related to (quantum) fluctuations of the superconducting
order parameter, both of its modulus and its phase. Such fluctuations may
-- and in general do -- involve $virtual$ states with the electron subsystem
driven {\it far from equilibrium}. Examples are quantum phase slips in
thin superconducting wires \cite{zgoz} and quantum tunneling of
vortices~\cite{tunne}. All such processes are also conveniently described
within the formalism developed below.

The analysis of the real time dynamics of a superconductor with strong
deviations of the quasiparticle distribution function from equilibrium
in general requires methods based on the Keldysh technique that keep track 
explicitly of the distribution function, see Ref.~\cite{noneq}. Strong
non-equilibrium real time dynamics is beyond the scope of the present
paper and we postpone the corresponding discussion to a forthcoming
publication~\cite{goz}. Here we would only like to point out that many
features of our imaginary time analysis can be directly generalized and
used also within the real time Keldysh technique. Thus, also for the real
time case a lot can be learned already from the present imaginary time
formulation of the problem.

The calculation that leads to the effective actions
Eqs.~(\ref{sintro},\ref{spha}) will be presented in the next Section.
Its physical content is discussed in Sections III (normal state) and
IV (superconducting state). Where possible, we give the explicit forms
of the propagators of the various fields in the hydrodynamic limit (in
Sections III and IV), whereas the more general (and more complicated)
expressions are deferred to Appendix B, where also the calculation of
polarization bubbles is outlined. The derivation of the Ward identities
is given in Appendix A.

Parts of the present paper were implicit in Refs.~\cite{avo,zgoz} and to
a somewhat lesser extent in Refs.~\cite{fluct,popov,klein,ambeg,aes,sz}.

\section{Model and Derivation of Effective Action}
The starting point for our analysis is a model Hamiltonian that includes a
short range attractive weak coupling BCS and a long range repulsive Coulomb
interaction. We represent the latter in terms of the fluctuating gauge
fields of electro-magnetism, $V$ and ${\bf A}$. The idea is to integrate
out the electronic degrees of freedom on the level of
the partition function, leaving us with an effective theory in terms
of collective fields\cite{popov,klein,ambeg,aes,sz}. The partition function
$Z$ is conveniently expressed as a path integral over the anticommuting
electronic fields $\bar{\psi}$, $\psi$ and the commuting gauge fields $V$
and ${\bf A}$, together with a gauge condition. The Euclidean action reads
\begin{eqnarray}\nonumber
	S=\int dx{\Big (}\bar{\psi}_{\sigma}
	[\partial_{\tau}-ieV+\xi({\bf \nabla}-\frac{ie}{c}{\bf A})]
	\psi_{\sigma}-\\
	-\lambda\bar{\psi}_{\uparrow}\bar{\psi}_{\downarrow}
	\psi_{\downarrow}\psi_{\uparrow}+ien_{i}V+
	[{\bf E}^{2}+{\bf B}^{2}]/8\pi{\Big )}\;.
\label{start}
\end{eqnarray}
Here $\xi({\bf \nabla})\equiv -{\bf \nabla}^{2}/2m-\mu$ describes a single
conduction band with quadratic dispersion, $\lambda$ is the BCS coupling
constant, $\sigma=\uparrow,\downarrow$ is the spin index, and $en_{i}$
denotes the background charge density of the ions. In our notation $dx$
denotes $d^{3}{\bf x}d\tau$ and we use units in which $\hbar$ and $k_{B}$
are set equal to unity. The field strengths are functions of the gauge
fields through ${\bf E}=-{\bf\nabla}V+(1/c)\partial_{\tau}{\bf A}$ and
${\bf B}={\bf\nabla}\times{\bf A}$ in the usual way for the imaginary
time formulation.

We use a Hubbard-Stratonovich transformation to decouple the BCS
interaction term and to introduce the superconducting energy gap
$\Delta=\vert\Delta\vert e^{i\varphi}$ as an order parameter
$$
	\exp\left(\lambda\int dx\bar{\psi}_{\uparrow}\bar{\psi}_{\downarrow}
		\psi_{\downarrow}\psi_{\uparrow}\right)=
	\left[\int{\cal D}^{2}\Delta e^{-\lambda^{-1}\int
		dx\mid\Delta\mid^{2}}\right]^{-1}
$$
\begin{equation}
	\int{\cal D}^{2}\Delta e^{-\int
		dx\left(\lambda^{-1}\mid\Delta\mid^{2}+
		\Delta\bar{\psi}_{\uparrow}\bar{\psi}_{\downarrow}+
		\bar{\Delta}\psi_{\downarrow}\psi_{\uparrow}
	\right)}\;,
\label{hubb-strat}
\end{equation}
where the first factor is for normalization and will not be important
in the following. As a result, the partition function now reads
$$
	Z=\int'{\cal D}^{2}\Delta{\cal D}V{\cal D}^{3}{\bf A}{\cal D}^{2}\Psi
	\exp\left(-S_{0}-\int dx\bar{\Psi}{\cal G}^{-1}\Psi\right)\;,
$$
\begin{equation}
	S_{0}[V,{\bf A},\Delta]=\int dx\left(\frac{{\bf E}^{2}+{\bf B}^{2}}
	{8\pi}+ien_{i}V+\frac{\vert\Delta\vert^{2}}{\lambda}\right)\;,
\label{part}
\end{equation}
where the prime on the integral denotes the restriction to a certain gauge
choice for the electromagnetic potentials $V$ and ${\bf A}$. Below in Section
III, we will sometimes use the Coulomb gauge ${\bf\nabla}\cdot{\bf A}\equiv 0$
in which the vector potential is completely transverse.
In Eq.~(\ref{part}) we have also introduced the Nambu spinor notation for
the electronic fields
\begin{equation}
	\Psi=\left(\begin{array}{c}\psi_{\uparrow}\\
		\bar{\psi}_{\downarrow}\end{array}\right)\;,\;\;
	\bar{\Psi}=\left(\begin{array}{cc}\bar{\psi}_{\uparrow} &
		\psi_{\downarrow}\end{array}\right)
\end{equation}
and the matrix Green's function in Nambu space
\begin{eqnarray}
	{\cal G}^{-1}&=&\left(\begin{array}{c}
	\partial_{\tau}-ieV+\xi({\bf \nabla}-\frac{ie}{c}{\bf A})
	\;\;\;\;\;\; \Delta\\ \noalign{\vskip 2 pt}
	\bar{\Delta} \;\;\;\;\;\;
	\partial_{\tau}+ieV-\xi({\bf \nabla}+\frac{ie}{c}{\bf A})
	\end{array}\right); \;\; \nonumber\\
	{\cal G}&=&\left(\begin{array}{c}{G} \;\;\;\; {F}\\
		\noalign{\vskip 2 pt} \bar{F} \;\;\;\; \bar{G}
		\end{array}\right),
\label{nambu}
\end{eqnarray}
with normal and anomalous Green's functions denoted by $G$ and $F$.

After a final Gaussian integration over the electronic degrees of freedom,
we are left with the effective action
\begin{eqnarray}
	S_{\rm eff}=-{\rm Tr}\ln{\cal G}^{-1}+S_{0}[V,{\bf A},\Delta]\;.
\label{seff0}
\end{eqnarray}
Here, the trace ``Tr'' denotes both a matrix trace in Nambu space and
a trace over internal coordinates or momenta and frequencies. In the
following ``tr'' is used to denote a trace over internal coordinates only.

\subsection{The Equations of Motion}
The Euler-Lagrange equations obtained by varying the action
Eq.~(\ref{seff0}) with respect to $V$ and ${\bf A}$ yield the two
Maxwell equations that describe Thomas-Fermi and London screening,
respectively. They read
\begin{eqnarray}\nonumber
	{\bf \nabla}\cdot{\bf E}&=&4\pi i e[n_{e}-n_{i}]\;,\\
	-\frac{1}{c}\partial_{\tau}{\bf E}+ {\bf \nabla}\times{\bf B}&=&
	\frac{4\pi}{c}{\bf j}_{e} \; .
\end{eqnarray}
Note that the ionic background contributes only to the charge density,
and not to the current, if we describe the system in the frame where
the ions are at rest.
Both the electronic density $n_{e}$ and current density ${\bf j}_{e}$ are
expressed through the diagonal elements $G$ and $\bar G$ of the matrix
(in Nambu space) electron Green's functions ${\cal G}$. Explicitly,
\begin{eqnarray}\nonumber
        n_{e}(x)&=&\mbox{Tr}[{\cal G}{\bf\sigma}_{3}]=\bar{G}(x,x)-{G}(x,x)\;,\\
	{\bf j}_{e}(x)&=&\frac{e}{m}\mbox{Tr}[(i{\bf\nabla}{\bf 1}+
	\frac{e}{c}{\bf A}{\bf \sigma}_{3}){\cal G}]\\\nonumber
	&=&\frac{e}{m}\left[(i{\bf \nabla}+\frac{e}{c}{\bf A}){ G}(x,y)+
	(i{\bf \nabla}-\frac{e}{c}{\bf A})\bar{ G}(x,y)\right]_{y=x}\!\!.
\end{eqnarray}
The matrices ${\bf \sigma}_{1,2,3}$ are the Pauli matrices and
below we will also use ${\bf \sigma}_{\pm}=\frac{1}{2}
({\bf \sigma}_{1}\pm i{\bf \sigma}_{2})$.

The electronic density $n_{e}$ is a function of the chemical potential
$\mu$, and in the presence of particle-hole asymmetry also of the energy
gap $\Delta$. At zero temperature it satisfies
$n_{e}(\mu=\epsilon_{F},\Delta=0)=n_{i}$.
In general the electronic density can be expanded as $n_{e}(\mu+ieV,\Delta)=
n_{i}+2ieN_{0}V+2N_{0}\Gamma\Delta^{2}+\cdots$, with the density of states
per spin $N_{0}$ and the particle-hole asymmetry parameter
$\Gamma=\partial_{\epsilon}N(\epsilon_{F})/(2\lambda N^{2}_{0})$, see
Ref.~\cite{charv}.
The requirement of overall charge neutrality makes the electrostatic
potential $V$ a function of the energy gap, $V_{\Delta}=
-i\Gamma\Delta^{2}/e$.
Longitudinal electric fields and deviations of the electronic density
$n_{e}$ from the ionic density $n_{i}$ are screened on the Thomas-Fermi
length scale $\lambda^{-2}_{TF}=8\pi e^{2}N_{0}$.
In the superconducting state, in addition the magnetic field ${\bf B}$ is
screened on the scale of the London penetration depth $\lambda^{-2}_{L}=4\pi
e^{2}n_{s}/mc^{2}$, where $n_{s}$ denotes the superfluid density.

Varying the action Eq.~(\ref{seff0}) with respect to $\bar{\Delta}$ yields
the BCS gap-equation for $\Delta$
\begin{equation}
        \Delta(x)=\mbox{Tr}[{\cal G}\sigma_{-}]=\lambda F(x,x)\;,
\end{equation}
with the anomalous Green's function $F$. The gap equation has a constant
solution $\Delta_{0}=\vert\Delta_{0}\vert\exp(i\varphi_{0})$ as well as
more complex time and space dependent solutions, such as vortices.

\subsection{Perturbation Expansion}
The effective action in Eq.~(\ref{seff0}) is the starting point for an
expansion around the constant saddle point solution $\Delta=\Delta_0$,
$V=V_{\Delta}$, and ${\bf A}=0$. We absorb the constant $V_{\Delta}$ in the
chemical potential $\mu$ from now on, so that it doesn't appear explicitly
in the following.

There are two ways of organizing the perturbation expansion. In this section
we will expand in $V$, ${\bf A}$, and $\Delta_{1}=\Delta-\Delta_{0}$, and to
this end split the inverse Green's function in Eq.~(\ref{nambu}) into an
unperturbed part ${\cal G}^{-1}_{0}$ and a perturbation ${\cal G}^{-1}_{1}$,
according to
\begin{eqnarray}
	\nonumber {\cal G}_{0}&=&\left(\begin{array}{c} G\;\;\;\; F \\
	\noalign{\vskip 2 pt}\bar{F}\;\;\;\;\bar{G}\end{array}\right)\;,\;\;
	{\cal G}^{-1}_{1}=\left(\begin{array}{cc}  K- L &\Delta_{1}\\
	\bar{\Delta}_{1} & - K- L \end{array}\right) \nonumber \;,\\
	K&=&\frac{m}{2}\left(\frac{e}{mc}\right)^{2}\!{\bf A}^{2}-ieV
	\; ,\;\; L=\frac{-i}{2}\frac{e}{mc}\{{\bf \nabla},{\bf A}\} \; ,
\label{pert1}
\end{eqnarray}
where $\{.,.\}$ denotes an anti-commutator. In the following it is
understood that the unperturbed Green's function has a chemical potential
$\mu+ieV_{\Delta}$ and an energy gap $\Delta_{0}$. Without loss of
generality, we choose $\Delta_{0}$ to be real.

At first sight the splitting we have just made seems not convenient and 
restricts severely the generality of our analysis, since we expand around
a state with constant phase of the order parameter.
However, this does not mean that we exclude, e.g., current carrying states
for which the phase of $\Delta$ depends on coordinates in an essential way
and is not small everywhere in the superconductor. We show below that the
actual parameters of our expansion are the gauge invariant linear
combinations of the electromagnetic potentials and the phase of the order
parameter, $\Phi=V-\dot{\varphi}/2e$ and ${\bf v}_{s}=\frac{1}{2m}
({\bf\nabla}\varphi-\frac{2e}{c}{\bf A})$. Only these parameters (and not
$V$, ${\bf A}$, and $\varphi$ separately) are required to be small within
the framework of our analysis. Thus, also states that carry a current
which is not necessary small can be described. This is a direct consequence
of gauge invariance which plays an important role in our consideration.

The other way of organizing the expansion is commented upon in subsection E.
It involves a unitary gauge transformation of the fields, after which one
expands directly in the gauge invariant fields $\Phi$ and ${\bf v}_{s}$, and
a manifestly real perturbation $\Delta_{1}=\vert\Delta_{1}\vert$. With the
help of a Ward-identity, the two expansions can be shown to be fully
equivalent. For pedagogical purposes we postpone the corresponding
discussion (see subsection E) and proceed with the expansion.

The trace of the inverse Green's function can be expanded in
${\cal G}^{-1}_{1}$ using
\begin{eqnarray}
	{\rm Tr}\ln{\cal G}^{-1}= {\rm Tr}\ln{\cal G}^{-1}_{0}+{\rm Tr}
	\sum^{\infty}_{n=1}\frac{(-1)^{n+1}}{n}
	({\cal G}_{0}{\cal G}^{-1}_{1})^{n} \; ,
\label{expand}
\end{eqnarray}
and only terms of order $n$ = 1 and 2 will be needed here.

The $n$ = 1 term in the effective action is $S_{1}=-{\rm Tr}({\cal G}_{0}
{\cal G}^{-1}_{1})$. The explicit evaluation of the trace yields
\begin{eqnarray}\label{first}
	S_{1}&=&-{\rm tr}[K(G-\bar{G})-L(G+\bar{G})+
		\bar{\Delta}_{1}F+\Delta_{1}\bar{F}]\\\nonumber
	&=&\int dx\left(-ien_{e}V+\frac{ne^2}{2mc^2}{\bf A}^2
	-\frac{\bar{\Delta}_{1}\Delta_{0}+\Delta_{1}\bar{\Delta}_{0}}
	{\lambda}\right)\;,
\end{eqnarray}
where we have used $G(x,x)=-\bar{G}(x,x)=-n_{e}/2$, and $F(x,x)=\Delta_{0} 
/\lambda$ according to the gap equation.

The second order contribution $S_{2}=\frac{1}{2}{\rm Tr}({\cal G}_{0}
{\cal G}^{-1}_{1})^{2}$ reads
\begin{eqnarray}\nonumber
	S_{2}=\frac{1}{2}{\rm tr}[GKGK+\bar{G}K\bar{G}K-2FK\bar{F}K
	  +2FL\bar{F}L\\
	+GLGL\!+\!\bar{G}L\bar{G}L\!-\!2GKGL\!+\!2\bar{G}K\bar{G}L
	  \!-\!2\bar{F}KFL\nonumber\\
	+2FK\bar{F}L+2F\bar{\Delta}_{1}GK-2\bar{F}\Delta_{1}\bar{G}K
	  +2G\Delta_{1}\bar{F}K\nonumber\\
	-2\bar{G}\bar{\Delta}_{1}FK-2F\bar{\Delta}_{1}GL
	  -2\bar{F}\Delta_{1}\bar{G}L-2G\Delta_1\bar{F}L\nonumber\\
	-2\bar{G}\bar{\Delta_1}FL+\bar{F}\Delta_1\bar{F}\Delta_1
	  \!+\!F\bar{\Delta}_{1}F\bar{\Delta}_{1}
	  \!+2G\Delta_{1}\bar{G}\bar{\Delta}_{1}].
\label{second}
\end{eqnarray}
As we are interested in contributions up to second order in the fields,
it suffices to take $K=-ieV$ in this expression.

\subsection{Longitudinal and Transverse Physics}
The next step is the evaluation of the traces in the expansion
Eq.~(\ref{second}). For two Green's functions $G$ and $G'$, and two fields
$A$ and $A'$, the following identities hold:
\begin{eqnarray}\nonumber
	{\rm tr}[GAG'A'] &=& \int{\,\mathchar'26\mkern-9mu{\!d}}
	q A(q)A'(-q)\{ 1\}_{GG'}\;;\\
	{\rm tr}[GAG'\{\nabla_{a},A'_{a}\}] &=& \nonumber\\
		=2i\int{\,\mathchar'26\mkern-9mu{\!d}}
	q A(q)A'_{a}\!\!\!\!&&\!\!\!\!(-q)
		\{(p+q/2)_{a}\}_{GG'} \label{trace}\;; \\ \nonumber
	{\rm tr}[G\{\nabla_{a},A_{a}\}G'\{\nabla_{b},A'_{b}\}] &=&\\
		=-4\int{\,\mathchar'26\mkern-9mu{\!d}} q A_{a}(q)A'_{b}(-q)
		\{(\!\!\!\!&&\!\!\!\!p+q/2)_{a}(p+q/2)_{b}\}_{GG'}\; ,\nonumber
\end{eqnarray}
where $a=x,y,z$ (repeated indices are summed over) and we have introduced
the short hand notation $q=({\bf q},\omega_{\mu})$,
$\int{\,\mathchar'26\mkern-9mu{\!d}} q\equiv
T\sum_{\omega_{\mu}}\int d^{3}{\bf q}/(2\pi)^{3}$, as well as the bracket
notation for polarization bubbles
\begin{eqnarray}
	\left\{B\right\}_{GG'}(q)=\int{\,\mathchar'26\mkern-9mu{\!d}}
	p B G(p+q)G'(p)\;.
\label{bracket}
\end{eqnarray}

Furthermore, we will split all fields in longitudinal and transverse
components with respect to the momentum ${\bf q}$ by making use of the
corresponding projection operators
\begin{eqnarray}
	P^{ab}_{L}=q^{a}q^{b}/{\bf q}^{2} \;,\;\;
	P^{ab}_{T}=\delta^{ab}-q^{a}q^{b}/{\bf q}^{2}
\end{eqnarray}
that satisfy $P^{2}_{i}=P_{i}$, $P_{L}P_{T}=P_{T}P_{L}=0$, and $P_{L}+P_{T}=1$.
With the help of the projection operators any vector field $V^{a}({\bf q})$
can be decomposed into a longitudinal part $V^{a}_{L}({\bf q})=
P^{ab}_{L}V^{b}({\bf q})= ({\bf V}\cdot{\bf q}/{\bf q}^{2})q^{a}$ and a
transverse part $V^{a}_{T}({\bf q})= P^{ab}_{T}V^{b}({\bf q})=
V^{a}-({\bf V}\cdot{\bf q}/{\bf q}^{2})q^{a}$.
Similarly, tensors $T^{ab}({\bf q})$ are decomposed into
$T^{ab}_{L}({\bf q})= {\rm Tr}[P_{L}T]P^{ab}_{L}$ and $T^{ab}_{T}({\bf q})=
\frac{1}{2}{\rm Tr}[P_{T}T]P^{ab}_{T}$, where the extra half in the latter
arises since ${\rm Tr}P_{T}=d-1=2$, in contrast to ${\rm Tr}P_{L}=1$.

We use these considerations to simplify the polarization bubbles that we
encounter in the perturbation expansion. By splitting $(p+q/2)_{a}$ into
longitudinal and transverse components, we decompose the bubbles as
\begin{eqnarray}\nonumber
	\left\{(p+q/2)_{a}\right\}_{GG'} &=&
	\left\{Q/{\bf q}^{2}\right\}_{GG'}q^{a}\;,\\ \nonumber
	\left\{(p+q/2)_{a}(p+q/2)_{b}\right\}_{GG'} &=&
	\left\{Q^{2}/{\bf q}^{2}\right\}_{GG'}P^{ab}_{L}+\\
	+\left\{({\bf p}\times{\bf q})^{2}/2{\bf q}^{2}\right\}_{GG'}
	\!\!\!\!&&\!\!\!\!P^{ab}_{T} \;,
\label{longtrans}
\end{eqnarray}
where in order to simplify notation, we have introduced $Q={\bf q}\cdot
({\bf p}+{\bf q}/2)$. In the following, we need only specific polarization
bubbles for which we introduce the notation
\begin{eqnarray}
	g_{0}=\left\{1\right\}_{GG}(q) \; &,& \;\; \label{thegs}
	g_{1}=\left\{Q/{\bf q}^{2}\right\}_{GG}(q) \;, \\ \nonumber
	g_{2}=\left\{Q^{2}/{\bf q}^{4}\right\}_{GG}(q) \; &,& \;\; g_{3}=
	\left\{({\bf p}\times{\bf q})^{2}/2{\bf q}^{4}\right\}_{GG}(q)\;.
\end{eqnarray}
Analogously the $f_{i}, h_{i}, k_{i}$ denote $\{..\}_{FF}$,
$\{..\}_{G\bar{G}}$, and $\{..\}_{FG}$ respectively.

Finally, as pointed out in Ref.~\cite{fluct}, it is advantageous to
also split the fluctuations of the energy gap in real longitudinal and
transverse components $\Delta_{1}=\Delta_{L}+i\Delta_{T}$. We will see
in the next subsection that $\Delta_{L}$ and $\Delta_{T}$ appear in the
effective action in rather different ways, related to their different
physical nature.

\subsection{The Effective Action}
The effective action up to second order in the fields $\Delta_{L}$,
$\Delta_{T}$, $V$, and ${\bf A}$ is found by gathering the terms from
$S_{0}$, $S_{1}$, and $S_{2}$ (Eqs.~(\ref{part},\ref{first},\ref{second})).
We split $S_{\rm eff}$ into a constant mean field part $S^{0}_{\rm eff}$
and a Gaussian fluctuation part $S^{2}_{\rm eff}$. No first order
contribution is present, as the terms $ien_{i}V$ and $-ien_{e}V$ cancel
by charge neutrality. We find
\begin{eqnarray}\label{sfin}
	S^{0}_{\rm eff}&=&-{\rm Tr}\ln{\cal G}^{-1}_{0}[\Delta_{0}]+
	\beta{\cal V}\Delta^{2}_{0}/\lambda \;,\\
	S^{2}_{\rm eff}&=&\int{\,\mathchar'26\mkern-9mu{\!d}} q{\Big [}
	  \frac{{\bf E}^{2}+{\bf B}^{2}}{8\pi}
	  +\frac{\Delta^{2}_{L}+\Delta^{2}_{T}}{\lambda}
	  +\frac{nm}{2}\left(\frac{e}{mc}\right)^{2}{\bf A}^{2} \nonumber\\
	&+&\left(\begin{array}{cccc} \Delta_{L} & \Delta_{T} &
		eV & \frac{eA^{a}}{mc} \end{array}\right)_{q}  \hat{\cal M}_{q}
	\left(\begin{array}{c}\Delta_{L} \\ \Delta_{T} \\ eV \\
		eA^{b}/mc \end{array}\right)_{-q}    {\Big ]}\;,\nonumber
\end{eqnarray}
where ${\cal V}$ denotes the volume of the system, $\beta$ the inverse
temperature, and we have introduced a matrix notation with
\begin{eqnarray}
	\hat{\cal M}_{q}&=&\left(\begin{array}{cccc}
	h_{0}+f_{0} & -ih_{0} & -2ik_{0} & -2q^{a}k_{1} \\
	ih_{0} & h_{0}-f_{0} & -2k_{0} & 2iq^{b}k_{1} \\
	-2ik_{0} & 2k_{0} & f_{0}-g_{0} & iq^{b}g_{1} \\
	-2q^{a}k_{1} & -2iq^{a}k_{1} & iq^{a}g_{1} & m^{ab}   \\
	\end{array}\right)\;, \nonumber \\
	m^{ab}&=&{\bf q}^{2}[(g_{3}+f_{3})P^{ab}_{T}
		+(g_{2}+f_{2})P^{ab}_{L}]\;.
\end{eqnarray}
In the above expression for the matrix $\hat{\cal M}_{q}$ it is understood
that all kernels are taken at momentum and frequency $q$.

The physical content of the effective action Eq.~(\ref{sfin}) can be brought
out more clearly by ``diagonalizing the matrix'', i.e., rewriting
Eq.~(\ref{sfin}) in terms of the eigenmodes. To this end we introduce the
superfluid velocity and the chemical potential for Cooper pairs in terms of
combinations of the gauge fields and the transverse gap fluctuations as
${\bf v}_{s}=\frac{1}{2m}[({\bf\nabla}\Delta_{T})/\Delta_{0}-\frac{2e}{c}
{\bf A}]$ and $\Phi=V-\frac{1}{2e}\dot{\Delta}_{T}/\Delta_{0}$.

The terms in Eq.~(\ref{sfin}) that couple the transverse phase-like gap
fluctuations to the other fields assume a diagonal form with the use of
the Ward-identity Eq.~(\ref{usefulf}). Little algebra shows explicitly that
\begin{eqnarray}\nonumber
	[\lambda^{-1}+h_{0}-f_{0}]\Delta^{2}_{T}+4k_{0}V(q)\Delta_{T}(-q)\\
	-(4ie/mc)k_{1}{\bf q}\cdot{\bf A}(q)\Delta_{T}(-q) \equiv
	(4e^{2}\Delta_{0}k_{0}/i\omega_{\mu})[\Phi^{2}-V^{2}]\nonumber\\
	+4m\Delta_{0}k_{1}P^{ab}_{L}\left[(e^{2}A^{a}A^{b}/m^{2}c^{2})
	-v^{a}_{s}v^{b}_{s}\right]\;,
\nonumber\end{eqnarray}
which we use to eliminate $\Delta_{T}$ from the action.
Furthermore, the field strengths are invariant under gauge transformations,
so that they may be expressed in $\Phi$ and ${\bf v}_{s}$ as
\begin{eqnarray}\nonumber
	\vert{\bf E}\vert^{2}&=&{\bf q}^{2}\vert\Phi(q)\vert^{2}+
		\frac{m^{2}\omega^{2}}{e^{2}}\vert{\bf v}_{s}(q)\vert^{2}+\\
	&&+\frac{m\omega}{e}[\Phi(q){\bf q}\cdot{\bf v}_{s}(-q)+
		\Phi(-q){\bf q}\cdot{\bf v}_{s}(q)]\nonumber\\
	\vert{\bf B}\vert^{2}&=&\frac{m^{2}c^{2}}{e^{2}}{\bf q}^{2}P^{ab}_{T}
		v^{a}_{s}(q)v^{b}_{s}(-q)\;,
\label{eandb}\end{eqnarray}
which allows us to rewrite the terms related to ${\bf A}$ and $V$ in terms
of ${\bf E}$ and ${\bf B}$. Finally, due to the Ward-identities
Eqs.~(\ref{usefulg1},\ref{usefulgQ}) the remaining terms in ${\bf A}$ and $V$
are seen to vanish.
Together with the propagator $\chi_{A}(q)=[2\lambda^{-1}+h_{0}(q)+h_{0}(-q)
+2f_{0}(q)]$ for the longitudinal gap fluctuations and introducing the
London and Josephson susceptibilities $\chi_{L}(q)= -8m\Delta_{0}k_{1}(q)$ and
$\chi_{J}(q)=8e^{2}\Delta_{0}k_{0}(q)/i\omega_{\mu}$, we obtain the ``normal''
and the ``superconducting'' contributions to the effective action
\begin{eqnarray}\nonumber
	S_{\rm sc}[\Delta_{L},\Phi,{\bf v}_s]&=&-{\rm Tr}\ln\left(
	  {\cal G}_{0}^{-1}[\Delta_0]{\cal G}_0[0]\right)+
	  \beta{\cal V}\Delta^{2}_{0}/\lambda+\\\label{Ssc}
	&&\!\!\!\!\!+\frac{1}{2}\int{\,\mathchar'26\mkern-9mu{\!d}}
	q\left(\chi_{A}\Delta^{2}_{L}+
	\chi_{J}\Phi^{2}+\chi_{L}{\bf v}_s^{2}\right),\\\nonumber
	S_{\rm nm}[{\bf E},{\bf B}]&=&-{\rm Tr}\ln{\cal G}_0^{-1}[0]+\\
	&&\!\!\!\!\!+\frac{1}{2}\int{\,\mathchar'26\mkern-9mu{\!d}}
	q\left(\chi_{E}{\bf E}^{2}
	-\chi_{M}{\bf B}^{2}\right)\;. \label{Snm}
\end{eqnarray}
Here we have also introduced the electric and magnetic susceptibilities
$\chi_{E}(q)=2e^{2}g_{1}(q)/(mi\omega_{\mu})$ and $\chi_{M}(q)=
2e^{2}[g_{2}(q)+f_{2}(q)-g_{3}(q)-f_{3}(q)]/(m^{2}c^{2})$.

The terms in Eq.~(\ref{sfin}) that couple the longitudinal gap fluctuations
to the other fields are nonvanishing only if particle-hole symmetry is broken.
Again by virtue of the Ward-identity Eq.~(\ref{usefulf}), they combine into
the action $S_{\rm ph}$ that describes the effects of particle-hole symmetry
breaking
\begin{eqnarray}\label{Sph}
	S_{\rm ph}&=&\int{\,\mathchar'26\mkern-9mu{\!d}} q{\Big (}
	\frac{1}{2}\chi_{\Gamma}(q)\Delta_{L}(q)\Delta_{L}(-q)+\\
	+&&\!\!\!\!\!\!\!\chi^{\parallel}_{\Gamma}(q)\Phi(q)\Delta_{L}(-q)
	+\chi^{\perp}_{\Gamma}(q){\bf q}\cdot{\bf v}_{s}(q)\Delta_{L}(-q)
	{\Big )}\;,\nonumber
\end{eqnarray}
where we have introduced the susceptibilities $\chi_{\Gamma}(q)=
h_{0}(q)-h_{0}(-q)$, $\chi^{\parallel}_{\Gamma}(q)=-2ie[k_{0}(q)+k_{0}(-q)]$,
and $\chi^{\perp}_{\Gamma}(q)=2[k_{1}(q)-k_{1}(-q)]$.

Defining
\begin{equation}
	S_{\rm em}[{\bf E},{\bf B}]=\int{\,\mathchar'26\mkern-9mu{\!d}}
	q\left(	\frac{{\bf E}^{2}+{\bf B}^{2}}{8\pi}\right)\;,
\label{Sem}
\end{equation}
we obtain the final result
$$
	S_{\rm eff}=S_{\rm sc}[\Delta_{L},\Phi,{\bf v}_s]
	+S_{\rm nm}[{\bf E},{\bf B}]
$$
\begin{equation}
	+S_{\rm ph}[\Delta_{L},\Phi,{\bf v}_s]
	+S_{\rm em}[{\bf E},{\bf B}]\;.
\label{sfinal}
\end{equation}
Using the standard definitions $\epsilon=1+4\pi\chi_{E}$ and
$\mu^{-1}=1-4\pi\chi_{M}$, we arrive at the form Eq.~(\ref{sintro}) quoted
in the introduction. The expressions Eq.~(\ref{Ssc}--\ref{sfinal}) represent
the main result of this Section, and a convenient starting point for any
study of dynamical processes.

Note that in arriving at Eq.~(\ref{sfinal}) the terms $\pm ien_{i/e}V$ from
$S_{0}$ and $S_{1}$ have cancelled, since on the average the electronic and
ionic charge densities cancel. This point was not appreciated in
Refs.~\cite{schak,aitch,stone,gait}, where no coupling to electromagnetism
was included and only the term from $S_{1}$ was found. 

We have decomposed the action into four parts: the superconducting
contribution $S_{\rm sc}$, the normal metallic contribution $S_{\rm nm}$,
the particle-hole symmetry breaking action $S_{\rm ph}$, and the action
of the free electromagnetic fields $S_{\rm em}$. Let us emphasize that the
possibility of such a decomposition is a direct consequence of the Ward
identities.
As these identities follow from gauge invariance only and do not depend on
the presence and concentration of impurities in a superconductor, we conclude
that the splitting of the full action into four parts in Eq.~(\ref{sfinal})
holds not only for clean superconductors but rather for an arbitrary
concentration of impurities.

\subsection{Gauge Invariance}
At this stage it is appropriate to discuss the consequences of gauge invariance
for the effective action Eq.~(\ref{sfinal}) a bit more deeply, see also
Appendix A. Inspecting Eq.~(\ref{sfinal}), we observe that the transverse
component of the energy gap $\Delta_{T}$ has completely disappeared from the
effective action. This is just the Anderson-Higgs mechanism~\cite{ah}: the
Goldstone mode $\Delta_{T}$ is ``gauged away'' and appears only within the
combinations $\Phi$ and ${\bf v}_{s}$. Since the electromagnetic field strengths
${\bf E}$ and ${\bf B}$ have been expressed in terms of $\Phi$ and ${\bf v}_{s}$,
the integral over the field $\Delta_{T}$ in the partition function factorizes
and contributes an irrelevant constant.

The partition function $Z$ can be represented in two equivalent ways. In the first,
the Goldstone mode is explicitly present and the four gauge field components are
restricted by a gauge condition. In the second, the Goldstone mode is ``eaten''
by the gauge condition and the four gauge field components are unrestricted.
Explicitly
\begin{eqnarray}\nonumber
	\int'{\cal D}\Delta_{L}{\cal D}\Delta_{T}{\cal D}V{\cal D}^{3}{\bf A}
	\exp(-S_{\rm eff}[\Delta_{L},\Delta_{T},V,{\bf A}])\equiv\\
	\equiv\int{\cal D}\Delta_{L}{\cal D}\Phi{\cal D}^{3}{\bf v}_{s}
	\exp(-S_{\rm eff}[\Delta_{L},\Phi,{\bf v}_{s}]),
\end{eqnarray}
where the prime on the first integral denotes that it is supplemented by a
gauge condition. In both cases 5 dynamical degrees of freedom are present.

We now return to the point raised in subsection B concerning the two possible
ways of organizing the expansion. Observe that the trace of the inverse Green's
function, which is the starting point of the perturbation expansion, is
invariant under unitary transformations
\begin{eqnarray}
	{\rm Tr}\ln{\cal G}^{-1}={\rm Tr}\ln{\cal U}{\cal G}^{-1}{\cal U}^{-1}\;.
\end{eqnarray}
Choosing ${\cal U}(\theta)=\exp(-i\sigma_{3}\theta/2)$ with an arbitrary space
and time dependent function $\theta(x)$, we obtain from the old Green's
function Eq.~(\ref{nambu}), the new inverse Green's function
$\tilde{\cal G}^{-1}=
\exp(-i\sigma_{3}\theta/2){\cal G}^{-1}\exp(i\sigma_{3}\theta/2)$ that reads
\begin{eqnarray}
	\tilde{\cal G}^{-1}=\left(\begin{array}{c}
	\partial_{\tau}-ie\Phi+\xi({\bf \nabla}+im{\bf v}_{s})
	\;\;\;\;\;\; e^{-i\theta}\Delta\\ \noalign{\vskip 2 pt}
	e^{i\theta}\Delta \;\;\;\;\;\
	\partial_{\tau}+ie\Phi-\xi({\bf \nabla}-im{\bf v}_{s})
	\end{array}\right)\;.
\label{green}
\end{eqnarray}
If $\Delta=|\Delta|e^{i\varphi}$, such a gauge transformation can be used to
make the energy gap in the Green's function real by choosing $\theta=\varphi$.
Instead, the superconducting phase appears in the chemical potential for Cooper
pairs $\Phi=V-\dot{\varphi}/2e$ that replaces $V$ and in the superfluid
velocity ${\bf v}_{s}=\frac{1}{2m}({\bf \nabla}\varphi-\frac{2e}{c}{\bf A})$
that replaces $-e{\bf A}/mc$. Thus, we may identify $\varphi$ with
$\Delta_{T}/\Delta_{0}$ from the previous subsection.

Note that the unitary operator ${\cal U}(\theta)$ is related to a local $U(1)$
gauge transformation
\begin{eqnarray}\nonumber
	V&\rightarrow& V-\frac{1}{2e}\dot{\theta} \; ; \;\; \\ \nonumber
	{\bf A}&\rightarrow& {\bf A}-\frac{c}{2e}{\bf \nabla}\theta \;;\\
	\nonumber \Psi=\left(\begin{array}{c}\psi_{\uparrow}\\
		\bar{\psi}_{\downarrow}\end{array}\right)
	&\rightarrow& {\cal U}(\theta) \Psi =
	\left(\begin{array}{c}e^{-i\theta/2}\psi_{\uparrow}\\
		e^{i\theta/2}\bar{\psi}_{\downarrow}\end{array}\right) \;;\\
	\Delta &\rightarrow& e^{-i\theta}\Delta\;,
\label{gauge}
\end{eqnarray}
that leaves the field strengths invariant. Since the phase of {\em all} charged
fields is rotated by a gauge transformation, one also should replace the term
$ien_{i}V$ in Eq.~(\ref{part}) by $ien_{i}\Phi$. The point being, that the ionic
background charge density is eventually made up by particles and corresponding
fields as well.

After the gauge transformation with $\theta=\varphi$, the perturbation expansion
can also be done on the level of $\tilde{\cal G}^{-1}$, in terms of the real
field $\Delta_{1}$ and
\begin{eqnarray}
	\tilde{K}=\frac{m}{2}{\bf v}^{2}_{s}-ie\Phi \;;\;\;
	\tilde{L}=\frac{i}{2}\{{\bf \nabla},{\bf v}_{s}\}\;.
\label{landk}
\end{eqnarray}
The whole derivation of the effective action is completely analogous and even
slightly easier. In this way, one again recovers the result Eq.~(\ref{sfinal}),
now without making use of the assumption about small electromagnetic
potentials and phase. These considerations emphasize the remarkable role of
gauge invariance and conclude our derivation of the effective action for a
BCS superconductor.

\section{The normal metal and Ginzburg-Landau theory}
\subsection{The normal metal}
As a first application of Eq.~(\ref{sfinal}), we will consider the normal
metal limit for temperatures above the critical temperature $T_{C}$.
If one puts $\Delta_{0}=0$ in the normal metal and discards fluctuations of the
energy gap, the electronic polarization terms can be expressed in terms of the
locally gauge invariant field strengths ${\bf E}$ and ${\bf B}$ only, as is
evident from Eq.~(\ref{sfinal}) since in the normal metal
$S_{\rm sc}\equiv 0$. This contrasts to the superconducting case where
also terms in $\Phi$ and ${\bf v}_{s}$ survive. For the normal metal we obtain
\begin{eqnarray}\nonumber
	S_{\rm nm}+S_{\rm em}&=&\int{\,\mathchar'26\mkern-9mu{\!d}} q{\Big (}
	\frac{{\bf E}^{2}+{\bf B}^{2}}{8\pi}
	-\frac{e^{2}}{{\bf q}^{2}}g_{0}{\bf E}^{2}+\\ &&\;\;\;\;\;\;\;\;
	+\frac{e^{2}}{m^{2}c^{2}}(g_{3}-g_{2}){\bf B}^{2}{\Big )}\;.
\label{semnm}
\end{eqnarray}
This action describes standard metal physics in the RPA approximation, as is
discussed for instance in Ref.~\cite{pn}. Its analysis is most conveniently
done in the Coulomb gauge. After an analytic continuation to real frequencies
$i\omega_{\mu}\rightarrow\omega+i\delta$ and $\vert\omega_{\mu}\vert\rightarrow
-i\omega$, the zeroes of the propagators describe modes of the electronic
system. Two useful limits are the clean and dirty limit, in which $\omega_{\mu},
D{\bf q}^{2}\gg\tau^{-1}_{r}$ and $\omega_{\mu}, D{\bf q}^{2}\ll\tau^{-1}_{r}$
respectively. Here $D=v^{2}_{F}\tau_{r}/3$ is the diffusion constant with
a single particle relaxation time due to impurity scattering $\tau_{r}$ and
Fermi velocity $v_{F}$.
For frequencies of the order of the Fermi energy or the plasma
frequency metallic systems are always in the clean limit.

In the dirty limit, $g_{0}$ is given by Eq.~(\ref{dirtynm}) and the part of the
action related to the longitudinal electric field has the form
\begin{eqnarray}
	S_{\rm nm}+S_{\rm em}=\int{\,\mathchar'26\mkern-9mu{\!d}} q{\Big (}1
	+\frac{8\pi e^{2}N_{0}D}{\vert\omega_{\mu}\vert+D{\bf q}^{2}}{\Big )}
	\frac{{\bf E}^{2}}{8\pi}\;.
\label{semnmel}
\end{eqnarray}
Since $8\pi e^{2}N_{0}= k^{2}_{TF}$, with Thomas-Fermi wavevector $k_{TF}$, the
low frequency part $\vert\omega_{\mu}\vert\ll D{\bf q}^{2}$ of this action
describes metallic screening. In the opposite high frequency limit the second
term of the action describes dissipation~\cite{caldl}: in terms of the
conductivity $\sigma=2e^{2}N_{0}D$, the second part reads
$\sigma{\bf E}^{2}/2\vert\omega_{\mu}\vert$.

In the clean high frequency limit $\omega_{\mu}\gg\tau^{-1}_{r},v_{F}q$,
the kernel $g_{0}$ is given by Eq.~(\ref{cleannm}), and using
$g_{0}\approx -N_{0}v^{2}_{F}q^{2} /(3\omega^{2})$, the longitudinal plasmon
at frequency $\omega^{2}_{p}=4\pi e^{2}n/m$ is recovered.

The remaining part of Eq.~(\ref{semnm}), related to the magnetic and transverse
electric fields, describes purely transverse physics. The kernel $g_{3}$ can be
approximated for low momenta and frequencies as $g_{3}=(p^{2}_{F}/3{\bf q}^{2})
g_{0}+N_{0}/12+\cdots$, and in the normal state the kernel $g_{2}$ can be
expressed through the Ward-identities Eq.~(\ref{usefulg1},\ref{usefulgQ}) as
$g_{2}=-nm/2{\bf q}^{2}+(m\omega/{\bf q}^{2})^{2}g_{0}$. In the dirty limit
the bubbles $g_{2}$ and $g_{3}$ almost cancel. The remaining action for the
transverse vector potential in the dirty limit is
\begin{eqnarray}
	S_{\rm nm}+S_{\rm em}=\int{\,\mathchar'26\mkern-9mu{\!d}} q{\Big (}
	q^{2}+\frac{\omega^{2}_{\mu}}{c^{2}}
	+\frac{4\pi\sigma\omega^{2}_{\mu}/c^{2}}
	{\vert\omega_{\mu}\vert+D{\bf q}^{2}}
	{\Big )}\frac{{\bf A}^{2}}{8\pi}\;.
\label{semnmma}
\end{eqnarray}
The low frequency limit describes the normal skin-effect, i.e., $i\omega=
c^{2}{\bf q}^{2}/(4\pi\sigma)$, for wavelengths larger than the mean free path
$l=v_{F}\tau_{r}$.

In the opposite clean limit and for small frequencies $\omega_{\mu}\ll v_{F}q$,
we find explicitly $g_{3}-g_{2}\approx N_{0}/12+\pi N_{0}
|\omega_{\mu}|/(4v_{F}q^{3})+\cdots$. The dispersion is now different,
$i\omega=4v_{F}\lambda^{2}_{L}(0)\vert q\vert^{3}/3\pi$, with the zero
temperature London length given by $\lambda^{-2}_{L}(0)=4\pi
e^{2}n/mc^{2}$, and is related to the anomalous skin-effect. Landau dia-magnetism
is recovered from the small constant term in $g_{3}-g_{2}=N_{0}/12+\cdots$, and
this is a $(v_{F}/c)^{2}$ correction to the ``1/8$\pi$'' in $S_{\rm em}$ in
Eq.~(\ref{Sem}): $\mu^{-1}=1+1/(4\lambda^{2}_{L}(0)p^{2}_{F})$.
Pauli paramagnetism is not present in the action Eq.~(\ref{sfinal}), since we
did not include a Zeeman coupling in the original model Eq.~(\ref{start}).
Finally, two transverse light modes with dispersion $\omega^{2}_{\mu}=
\omega^{2}_{p}+c^{2}{\bf q}^{2}$ are present in the high frequency clean limit.

\subsection{The Ginzburg-Landau Expansion}
In the normal metal close to $T_{C}$, fluctuations of the order parameter
$\Delta$ can be studied using Ginzburg-Landau theory~\cite{gl,gorko}, or its
time dependent (TDGL) generalization~\cite{schmi,fukuy,gorel,ambsc}. Within
our present formalism, the TDGL effective action is readily derived starting
from the action Eq.~(\ref{sfinal}). The expansion in $\Phi$ and ${\bf v}_s$ is
already performed and we need only to expand all terms in powers of $\Delta$.
In addition, we need parts of the third and fourth order terms in the expansion
of the inverse Green's function Eq.~(\ref{expand}). They are calculated using
the normal metal Green's functions with zero energy gap, see App.~B3. This
procedure is quite standard~\cite{gorko} (see also
Eqs.~(\ref{cleannm},\ref{dirtynm})). The superconducting part of the action
takes the form
\begin{eqnarray}\label{Ssgl}
	S_{\rm sc}&=&N_{0}\frac{1}{\beta}\sum_{\omega_\mu}\int d{\bf x}^{3}
	{\Big (}\frac{\pi |\omega_\mu|}{8T}|\Delta|^2{\Big )}\\\nonumber
	&+&\!N_{0}\int dx{\Big (}
	\bar{\Delta}{\Big [}\ln\left(\frac{T}{T_{C}}\right)
	-\xi^{2}(0){\bf\nabla}^{2}+4m^{2}\xi^{2}(0){\bf v}^{2}_{s}\\
	&&\;\;\;\;\;\;\;\;\;\;\;+\Gamma(\partial_{\tau}-i2eV)
	+2e^{2}b\Phi^{2}{\Big ]}\Delta
	+\frac{b}{2}|\Delta|^{4}{\Big )}\;.\nonumber
\end{eqnarray}
Here we have introduced the coherence length $\xi^2(0)=\frac{\pi}{8T}D$ in
the dirty limit and $\xi^2(0)=(7\zeta(3)/48\pi^2)(v_F/T)^2$ in the clean
limit. The coefficient $b=7\zeta(3)/(8\pi^{2}T^{2})$ is derived in
Eq.~(\ref{4bubb}). The small coefficient $\Gamma=N^{\prime}_{0}/2\lambda
N^{2}_{0}$ arises due to particle hole asymmetry~\cite{fukuy} and is usually
neglected. The term containing this coefficient describes the small
difference between electronic and ionic densities resulting from the
fluctuations of the order parameter. It occurs in the quoted gauge
invariant form, since the time derivative from the Cooperon combines
with the term $\sim\Delta^{2}V$ from the third order expansion. We will
see in the next Section that the same term arises also at low temperatures.

The term involving the second order space derivative can be combined with
the ${\bf v}^{2}_{s}\Delta^{2}$ term from the fourth order terms in the
expansion into one gauge invariant second order derivative
$\xi^{2}(0)\vert({\bf\nabla}-i\frac{2e}{c}{\bf A})\Delta\vert^{2}$.
The $\Phi^{2}\Delta^{2}$ term, however, does not straightforwardly combine
with time derivatives into gauge invariant time derivatives. The reason is
that close to $T_{C}$ dissipative and Hamiltonian frequency dependences mix.
As an example, the dissipative $|\omega_\mu|$ term~\cite{caldl} in
Eq.~(\ref{Ssgl}), which turns into the dissipative time derivative of the
real time TDGL equation after an analytic continuation, clearly cannot be
made gauge invariant. Since the second and higher order time derivatives
are usually irrelevant as compared to the dissipative $|\omega_\mu|$ term,
we do not include them in Eq.~(\ref{Ssgl}).

Let us also note that the expression Eq.~(\ref{Ssgl}) is correct only in the
limit of low frequencies and wave vectors $\omega_{\mu},Dq^{2} <4\pi T$. For
larger frequencies the expression becomes more complicated, as we can no longer
expand the kernel $h_{0}$ in $\omega_{\mu}/4\pi T$, e.g.,
$\Psi(1/2+\omega_{\mu}/4\pi T)-\Psi(1/2)\to \pi\omega_{\mu}/8T$ ($\Psi$ is
the digamma function, see App.~B).
Also the gradient terms in Eq.~(\ref{Ssgl}) should be modified in this case.
As the corresponding expressions turn out to be quite tedious we do not present
them here. For some problems, however, these modifications become significant,
especially because the validity of the GL expansion Eq.~(\ref{Ssgl}) is
restricted to temperatures $T\sim T_{C}$, in which case $\omega_\mu$ is never
really smaller than $4\pi T$.

The action $S_{\rm nm}$ is also important for the description of the dynamical
properties of the superconductor and describes dissipation in the ``sea'' of
the remaining normal electrons. It explicitly depends on the order parameter,
since also the polarization bubbles can be expanded in $\Delta$, giving
rise to additional contributions. Expanding the bubble $g_{0}+f_{0}$ (see
Eqs.~(\ref{dirtyf},\ref{dirtyghk})) up to the second order in $\Delta$, we
obtain in the limit of small frequencies and wave vectors~\cite{maki,thomp}
\begin{equation}
	g_{0}+f_{0}=-N_{0}\frac{D{\bf q}^2}{|\omega_\mu|}-
	N_0\frac{\pi\Delta^2}{4|\omega_\mu|T}\;.
\end{equation}
For the part of the effective action concerned with the electric field, we find
\begin{eqnarray}
	S_{\rm nm}+S_{\rm em}=\int{\,\mathchar'26\mkern-9mu{\!d}}
	q{\Big(}1+\frac{4\pi\sigma}
	{|\omega_\mu|}+\frac{2\pi^2 e^2 N_{0}}{{\bf q}^2|\omega_\mu|T}
	\Delta^2{\Big )}\frac{{\bf E}^{2}}{8\pi}\;.
\label{SNgl}\end{eqnarray}
The dependence of the action $S_{\rm nm}$ on $\Delta$ is important in dirty
superconductors and accounts for the Maki-Thompson fluctuation enhancement
of the conductivity near $T_{C}$~\cite{maki,thomp}.
The Azlamazov-Larkin fluctuation correction to the conductivity is already
present in the TDGL action $S_{\rm sc}$ Eq.~(\ref{Ssgl}) due to the presence
of the ${\bf v}^{2}_{s}\Delta^{2}$ term~\cite{al}.

Extensions of the elegant TDGL action into the superconducting phase have
turned out to be hard~\cite{abrts}. Only in the presence of a
large amount of paramagnetic impurities~\cite{gorel} or very close to
$T_{C}$~\cite{ambsc} is this possible.

\section{Dynamics at Lower Temperatures}
\subsection{Electromagnetism}
We will now evaluate the contents of Eq.~(\ref{sfinal}) in the superconducting
state with an equilibrium gap $\Delta_{0}$. We first focus on the parts
involving ${\bf E}$ and $\Phi$ that are related to the electric field and
the Josephson relation. For the prefactor of $\Phi^{2}$, we take the bubble
$4\Delta_{0}k_{0}/i\omega_{\mu}=N_{0}n_{s}/n$ at zero frequency and momentum.
At small frequency and momentum the bubble that multiplies ${\bf E}^{2}$ is 
$g_{1}/(mi\omega_{\mu})=N_{0}n_{n}/(nq^{2})$, see Appendix B. The action reads
\begin{eqnarray}
	S_{\rm eff}=\int{\,\mathchar'26\mkern-9mu{\!d}}
	q{\Big (}\left[\frac{1}{8\pi}
	+\frac{e^{2}N_{0}}{q^{2}}\frac{n_{n}}{n}\right]{\bf E}^{2}
	+e^{2}N_{0}\frac{n_{s}}{n}\Phi^{2} {\Big )}\;.
\label{semscma}
\end{eqnarray}
It describes metallic screening of the electrostatic potential $V$ with the
full electronic density $n$, to which both terms in Eq.~(\ref{semscma})
contribute, and superconducting screening with superfluid density $n_{s}$,
only through the second term, to enforce the Josephson relation
$V=\dot{\varphi}/2e$.
At higher frequencies and momenta, weight shifts from the term $\sim n_{s}$
to the term $\sim n_{n}$ in such a way that the plasma frequency and
Thomas-Fermi screening length remain constant.
The higher order frequency and momentum dependence of the kernels is not easy
to extract. For low frequency and momentum, one typically finds corrections
of order $\omega_{\mu}/\Delta_{0}$ and $v_{F}q/\Delta_{0}$ or $Dq^{2}/\Delta_{0}$.
In the opposite limit, for high frequencies and momenta, the kernels reduce
to their normal state form, and we recover dissipation in the dirty limit for
frequencies $\Delta_{0}\lesssim\omega_{\mu}\lesssim \tau^{-1}_{r}$ (cf.
Eq.~(\ref{semnmel})). In the clean limit such an intermediate frequency regime
does not exist.

We now turn to the parts of Eq.~(\ref{sfinal}) that are related to the magnetic
field and the superfluid velocity. To lowest order in the external momentum
$g_{3}\approx g_{2}\approx (p^{2}_{F}/3q^{2})g_{0}+\cdots$ and $f_{3}\approx
f_{2}\approx (p^{2}_{F}/3q^{2})f_{0}+\cdots$, so that the combination
$g_{3}+f_{3}-g_{2}-f_{2}$ is equal to zero in the $q\rightarrow 0$ limit.
What remains is the evaluation of $-4m\Delta_{0}k_{1}= mn_{s}/2$, and we obtain
\begin{eqnarray}
	S_{\rm eff}=\int{\,\mathchar'26\mkern-9mu{\!d}}
	q{\Big (}\frac{{\bf B}^{2}}{8\pi}+
	\frac{mn_{s}}{2}v^{2}_{s}{\Big )} \;.
\label{sb}
\end{eqnarray}
This action describes transverse screening of the magnetic field in a
superconductor and is related to the London theory.

Summarizing: at high frequencies $\omega_{\mu}\gg\Delta_{0}$ and momenta
$q\gg\xi^{-1}$ the electromagnetic properties are those of a normal metal.
At low frequencies and momenta a superconductor screens, in addition to
electric fields, also magnetic fields.

\subsection{Dynamics of the Energy Gap}
It is clear from Eq.~(\ref{Ssc}) that the dynamics of the fluctuations of the
amplitude of the energy gap $\Delta_{L}$ are governed by the combination
$\chi_{A}=[2\lambda^{-1}+2f_{0}(q)+h_{0}(q)+h_{0}(-q)]$. In the clean
limit we obtain to lowest order in $\omega_{\mu}/\Delta_{0}$ and
$v_{F}q/\Delta_{0}$ at zero temperature (see Appendix B)
\begin{eqnarray}
	S_{\rm sc}=N_{0}\int{\,\mathchar'26\mkern-9mu{\!d}}
	q{\Big (}1+\frac{1}{12}	\frac{\omega^{2}_{\mu}}{\Delta^{2}_{0}}+
	\frac{v^{2}_{F}{\bf q}^{2}}{36\Delta^{2}_{0}}{\Big )}\Delta^{2}_{L} \;.
\label{sgap}
\end{eqnarray}
This result is often quoted in the literature as evidence for the existence
of a TDGL-like theory with a second order time derivative at zero
temperature~\cite{aitch}.
Its use is restricted, however. The mode that it describes has dispersion
$\omega^{2}=12\Delta^{2}_{0}+(1/3)v^{2}_{F}{\bf q}^{2}$, i.e., it is gapped
to an energy where the expansion in $\omega_{\mu}/\Delta_{0}$ does not make
sense. The correct mode of the amplitude of the energy gap is heavily
overdamped due to the coupling to particle-hole pairs, and starts off at
frequencies $2\Delta_{0}$ as discussed in Refs.~\cite{popov,klein}.
For driven situations at small frequency and momenta on the other hand,
the expansion in Eq.~(\ref{sgap}) is useful.
At nonzero temperatures, the exact frequency and momentum dependence of
$\chi_{A}$ is rather involved unfortunately.

\subsection{Particle-hole asymmetry}
We now turn to the particle-hole symmetry breaking action Eq.~(\ref{Sph}).
To lowest order in the frequency and momentum, we obtain $\chi_{\Gamma}(q)=
h_{0}(q)-h_{0}(-q)=2N_{0}\Gamma i\omega_{\mu}$, $\chi^{\parallel}_{\Gamma}(q)=
-2ie[k_{0}(q)+k_{0}(-q)]=-4ieN_{0}\Gamma\Delta_{0}$, and
$\chi^{\perp}_{\Gamma}(q)=2[k_{1}(q)-k_{1}(-q)]=0$. Together with the term
$|\Delta_{1}|^{2}V$ from the third order expansion of the effective action,
we find
\begin{eqnarray}
	S_{\rm ph}=-2ieN_{0}\Gamma\int dx \; \Phi\left(\vert\Delta_{0}
	+\Delta_{1}\vert^{2}-\vert\Delta_{0}\vert^{2} \right)\;,
\label{sgap+}
\end{eqnarray}
as announced in the introduction. Apparently, this contribution to the
action is independent of temperature and the mean free path. Close to
$T_{C}$ exactly the same term, with $\Delta_{0}\equiv 0$ appeared already
in the TDGL expansion (see Eq.~(\ref{Ssgl})). The coupling constant
$\Gamma$ of this term is usually small, and $S_{\rm ph}$ is irrelevant,
except for inhomogeneous problems related to vortex motion~\cite{avo}.

In the core of a vortex the energy gap goes to zero, and this local variation
of the gap induces a local charge density modulation, for which the action
$S_{\rm ph}$ contains the source term~\cite{charv}. Moreover, due to the
singular phase field around a vortex, together with the suppression of the
gap in the core, the action $S_{\rm ph}$ gives rise to a small additional
force per length on vortices, $-2\pi N_{0}\Gamma\Delta^{2}_{0}{\bf v}_{L}
\times\hat{\bf z}$, which is proportional and perpendicular to the vortex
velocity ${\bf v}_{L}$ ($\hat{\bf z}$ is the unit vector along a vortex in
the direction of the magnetic field)~\cite{avo,dors}. We do not find evidence
for the much larger force $-\pi n_{s}{\bf v}_{L}\times\hat{\bf z}$ found in
Ref.~\cite{aoth}.

Note that this ``topological'' force is only one contribution to the several
different forces on a vortex; for a review of the other forces on
a vortex-line, see, e.g., Ref.~\cite{sonin}.

\subsection{The Uncharged Limit}
The limit where the electronic charge vanishes has received some attention
recently~\cite{greit,schak,aitch,stone}. Although this case is realized in
superfluid $^3$He, the different order parameter symmetry makes any s-wave
considerations less useful. Furthermore, the interactions between uncharged
$^3$He atoms is very different from the electron-electron interactions as
described by our starting point Eq.~(\ref{start}). In particular, $^3$He
atoms are neutral and in the $^3$He system no background charge is present.

For completeness, however, we also discuss the uncharged limit of our model
Eq.~(\ref{start}).
Putting $e\rightarrow 0$ in Eqs.~(\ref{Ssc},\ref{Sph}), we find to lowest
order in momentum and frequency for the phase part of the action
\begin{eqnarray}
	S_{\rm sc}\!+\!S_{\rm ph}\!=\!N_{0}\!\!\int\!\!
	{\,\mathchar'26\mkern-9mu{\!d}} q{\Big (}\!\!
	-\!i\Gamma\Delta^{2}_{0}\dot{\varphi}\!+\!
	\frac{n_{s}}{4n}{\Big [}\dot{\varphi}^{2}\!
	+\!\frac{v^{2}_{F}}{3}({\bf \nabla}\varphi)^{2}{\Big ]}{\Big )},
\label{sphase}
\end{eqnarray}
which gives the standard acoustic Bogoliubov-Anderson mode with velocity
$v=v_{F}/\sqrt{3}$. The velocity is temperature independent, and non-critical
even around $T_{C}$. Note that the phase action Eq.~(\ref{sphase}) is
different from the ones obtained in Refs.~\cite{greit,schak,aitch,stone}.
In particular, in contrast to Refs.~\cite{greit,schak,aitch,stone} no
large topological term $in_{e}\dot{\varphi}/2$ is present, only a much smaller
term proportional to the particle-hole asymmetry. The difference
can be traced to the near perfect cancellation between ionic and electronic
charge densities in our case.

For superconducting Bose-liquids a large topological term is present in the
effective action, since the bosonic field itself can be taken as the order
parameter at low temperatures and as a result the phase is dual to the density.
However, for Fermionic superconductors one rather expects that the phase of
the order parameter is dual to the amplitude of the order parameter, i.e., to
the energy gap. As a consequence, instead of a term $in_{e}\dot{\varphi}$, we
expect a term proportional to $i\Delta^{2}\dot{\varphi}$ in the effective
action.
This is just the content of Eqs.~(\ref{sgap+},\ref{sphase}), which shows that
the constant of proportionality is given by the particle-hole asymmetry
parameter $\Gamma$.

\section{Conclusion}
We have reviewed the derivation of the effective theories for BCS
superconductors and discussed the corresponding dynamics of electromagnetism
and the amplitude of the energy gap. Our main result Eq.~(\ref{sfinal}) is
a good starting point for investigations of quantum dynamical and statistical
problems in BCS superconductivity. 

We have stressed the role of gauge invariance and the corresponding Ward
identities that express particle number conservation. Although a perturbation
expansion can violate gauge invariance, the Ward identities allowed us to obtain
explicitly gauge invariant results. In particular, we have demonstrated how
the Anderson-Higgs mechanism occurs within BCS theory. In contrast, the role
of Galilei invariance that was stressed in Refs.~\cite{greit,schak,aitch,stone}
does not seem to play an important role in real BCS materials.

Furthermore, we included the effect of particle-hole asymmetry in our
considerations. We find a small topological term proportional to the
particle-hole asymmetry, that leads to an additional Hall-force on vortices
(apart from the Kopnin-Kravtsov, Magnus, and Iordanksii forces~\cite{sonin})
as discussed in Refs.~\cite{avo,volo}.
Also, we have seen that the structure of the theory is essentially the same
in the clean and dirty limits. In particular, the prefactor of the topological
term does not depend on the electronic mean free path.
The main difference between the clean and dirty limits is the presence of an
intermediate dissipative regime $\Delta_{0}\lesssim \omega_{\mu}, v_{F}q\lesssim
\tau^{-1}_{r}$ in the dirty limit. This difference shows up for instance in the
quantum dynamics of low dimensional superconductors~\cite{zgoz}.

{\bf Acknowlegdement} We thank H. Katzgraber, G. Sch\"{o}n, U. Eckern,
R. Fazio, D. Geshkenbein, D. Rainer, K-H. Wagenblast, and G. Zimanyi for
discussions on several aspects of our results. The support by the Swiss
National Foundation, the Deutsche Forschungsgemeinschaft within SFB 195,
and NSF Grant 95-28535 is gratefully acknowledged. One of us (D.S.G.) also
acknowledges partial support from the International Centre for Fundamental
Physics in Moscow.

\appendix
\section{The Ward Identity}
An important step is the derivation of the Ward identity related to gauge
invariance. On the level of vertex functions this Ward identity is discussed
for instance in Ref.~\cite{respo}. Here we will derive the Ward identity on the
level of the Green's function in order to obtain relations between the
polarization bubbles. Our derivation holds for an arbitrary concentration
of impurities in a superconductor.

In the clean limit the Ward identity also holds as an algebraic identity. For
the case of normal electrons, with single particle Green's function
$G(p)=[i\omega_{\nu}+\xi_{\bf p}]^{-1}$, it is
\begin{eqnarray}
	G(p)-G(p+q)=(i\omega_{\mu}+Q/m)G(p)G(p+q)\;,
\label{wime}\end{eqnarray}
where $Q/m=\xi({\bf p}+{\bf q})-\xi({\bf p})$.

A general way of establishing the Ward identity is to consider the change in
the Green's function upon rotating the electronic phase by $\varphi$.
We have on the one hand (we expand in $\varphi$)
\begin{eqnarray}\nonumber
	{\cal G}_{\phi}(x,x')&=&e^{i\phi(x)\sigma_{3}/2}{\cal G}(x,x')
	e^{-i\phi(x')\sigma_{3}/2}\\ \nonumber
	&=&{\cal G}(x,x')+\delta{\cal G}(x,x')
\end{eqnarray}
so that
\begin{eqnarray}\label{wi+}
	\delta{\cal G}(x,x')&=&\frac{i}{2}[\phi(x)\sigma_{3}
	{\cal G}(x,x')-{\cal G}(x,x')\sigma_{3}\phi(x')]\\\nonumber
	&=&\frac{i}{2}\int{\,\mathchar'26\mkern-9mu{\!d}}
	q{\,\mathchar'26\mkern-9mu{\!d}} p e^{iqx}e^{ip(x-x')}
	\phi_{q}[\sigma_{3}{\cal G}_{p}-{\cal G}_{p+q}\sigma_{3}]\;,
\end{eqnarray}
whereas on the other hand
\begin{eqnarray}\nonumber
	{\cal G}^{-1}_{\phi}(x,x')&=&e^{i\phi(x)\sigma_{3}/2}
	{\cal G}^{-1}(x,x')e^{-i\phi(x')\sigma_{3}/2}\\ \nonumber
	&=&{\cal G}^{-1}(x,x')+\delta{\cal G}^{-1}(x,x')
\end{eqnarray}
so that
\begin{eqnarray}\nonumber
	\delta{\cal G}^{-1}(y,y')&=&\delta(y-y'){\Big (}
	-\frac{i}{2}\dot{\phi}\sigma_{3}+
	\frac{i}{4}\{{\bf \nabla},{\bf \nabla}\phi\}\hat{1}+\\ \nonumber
	&&+i\phi\Delta\sigma_{+}-i\phi\bar{\Delta}\sigma_{-} 
	{\Big )}
\end{eqnarray}
and
\begin{eqnarray}\nonumber
	\delta{\cal G}(x,x')&=&
	-\int dydy'{\cal G}(x,y)\delta{\cal G}^{-1}(y,y'){\cal G}(y',x')\\
	\label{wi-}
	&=&-\int{\,\mathchar'26\mkern-9mu{\!d}} q{\,\mathchar'26\mkern-9mu{\!d}}
	p e^{iqx}e^{ip(x-x')}\phi_{q}{\cal G}_{p+q}\\
	&&\left[\frac{\omega_{\mu}}{2}\sigma_{3}-\frac{i}{2}\frac{Q}{m}\hat{1}+
	i(\Delta\sigma_{+}-\bar{\Delta}\sigma_{-})\right]{\cal G}_{p} \; .
\nonumber\end{eqnarray}
Comparison of Eqs.~(\ref{wi+},\ref{wi-}) leads to
\begin{eqnarray}\nonumber
	\sigma_{3}{\cal G}_{p}-{\cal G}_{p+q}\sigma_{3}=\\
	={\cal G}_{p+q}	\left[i\omega_{\mu}\sigma_{3}+Q/m\hat{1}-
	\Delta\sigma_{+}+\bar{\Delta}\sigma_{-}\right]{\cal G}_{p}\;.
\label{wisc}\end{eqnarray}
The restriction to the normal metal yields again Eq.~(\ref{wime}). In the
superconducting case the Ward-identities are slightly more complicated. The
upper left and upper right components of Eq.~(\ref{wisc}) read explicitly
\begin{eqnarray}\nonumber
	G(p)-G(p+q)=(i\omega_{\mu}+Q/m)G(p)G(p+q)+\\\nonumber
	+(-i\omega_{\mu}+Q/m)\bar{F}(p)F(p+q)-\\\label{wiexg}
	-2\Delta\bar{F}(p)G(p+q)+2\bar{\Delta}G(p)F(p+q)\;,\\\nonumber
	F(p)+F(p+q)=(i\omega_{\mu}+Q/m)F(p)G(p+q)+\\\nonumber
	+(-i\omega_{\mu}+Q/m)\bar{G}(p)F(p+q)-\\
	-2\Delta\bar{G}(p)G(p+q)+2\bar{\Delta}F(p)F(p+q)\;.
\label{wiexf}
\end{eqnarray}
These identities can be used to generate the Ward identities for the
electronic polarization bubbles, by tracing them over the internal momentum
and frequency $p$ together with some function. The trace of Eq.~(\ref{wiexg})
with $1$ and $Q/m$ immediately gives
\begin{eqnarray}\label{usefulg1}\nonumber
	0&=&i\omega_{\mu}[g_{0}(q)-f_{0}(q)]+{\bf q}^{2}/m[g_{1}(q)+f_{1}(q)]\\
	&&\hspace*{2cm}+2\Delta_{0}[k_{0}(q)-k_{0}(-q)]\\\nonumber
	-n/2&=&i\omega_{\mu}[g_{1}(q)-f_{1}(q)]+{\bf q}^{2}/m[g_{2}(q)+f_{2}(q)]\\
	&&\hspace*{2cm}+2\Delta_{0}[k_{1}(q)+k_{1}(-q)]\;,
\label{usefulgQ}
\end{eqnarray}
and the trace of Eq.~(\ref{wiexf}) with $1$ yields
\begin{eqnarray}\nonumber
	\Delta_{0}\lambda^{-1}&=&\Delta_{0}[f_{0}(q)-h_{0}(q)]+
	i\omega_{\mu}k_{0}(-q)\\&&\hspace*{1cm}-({\bf q}^{2}/m)k_{1}(-q)\;.
\label{usefulf}
\end{eqnarray}
The last three Eqs.~(\ref{usefulg1},\ref{usefulgQ},\ref{usefulf}) are used in
the text in Section 2. They are a result of gauge-invariance (particle number
conservation) and hold also after an impurity averaging procedure.
The identities can be simplified further using $f_{1}=0$ which holds by
symmetry.

The previous discussion was based on the gauge symmetry that is generated by
$\exp(i\sigma_{3}\phi/2)$. Along completely analogous lines, the invariance with
respect to rotations by $\exp(i{\bf 1}\phi/2)$ leads to
\begin{eqnarray}\nonumber
	i\omega_{\mu}f_{0}(q)&=&\Delta_{0}[k_{0}(q)-k_{0}(-q)]\;,\\
	-({\bf q}^{2}/m)f_{2}(q)&=&\Delta_{0}[k_{1}(q)+k_{1}(-q)]\;.
\label{usefulk}
\end{eqnarray}
These last two identities are not important in the derivation of the effective
action Eq.~(\ref{sfinal}). They do, however, reduce the amount of work needed to
explicitly evaluate all the different polarization bubbles.

\section{The Polarization Bubbles}
In this Appendix the polarization bubbles that are used in the text are
discussed and results in several limits are summarized.

\subsection{Clean Limit}
We evaluate the kernels by doing the sum over Matsubara frequencies first. The
notation $E=\sqrt{\xi^{2}_{\bf p}+\Delta^{2}_{0}}$ and $E'= 
\sqrt{\xi^{2}_{{\bf p}+{\bf q}}+\Delta^{2}_{0}}$ is used, as well as
$\int{\,\mathchar'26\mkern-9mu{\!d}}\Omega$ to denote a normalized angular
integration. In our notation
the unperturbed Green's function in momentum space reads explicitly
\begin{eqnarray}
	\left(\begin{array}{c}G\;\;F\\ \bar{F}\;\;\bar{G}\end{array}\right)=
	\frac{1}{\omega^{2}_{\nu}+\xi_{\bf p}^{2}+\Delta^{2}_{0}}
	\left(\begin{array}{c}-i\omega_{\nu}+\xi_{\bf p}\;\;\;\;\;\Delta_{0}\\
	\bar{\Delta}_{0}\;\;\;\;-i\omega_{\nu}-\xi_{\bf p}\end{array}\right).
\label{unpert}
\end{eqnarray}
For the bubble $f_{0}$ we obtain by standard contour integration and
ordering terms
\begin{eqnarray}\nonumber
	f_{0}(q)&=&\int\frac{d^{3}{\bf p}}{(2\pi)^{3}}\frac{1}{\beta}
	\sum_{\omega_{\nu}}F({\bf p}+{\bf q},\omega_{\nu}+\omega_{\mu})
	F({\bf p},\omega_{\nu})\\\nonumber
	&=&\int d\xi N(\xi)\int{\,\mathchar'26\mkern-9mu{\!d}}\Omega
	\frac{1}{2EE'}{\Big (}
	\frac{[1-f_{E'}-f_{E}]}{\omega^{2}_{\mu}+(E'+E)^{2}}S_{f}+\\
	&&\hspace*{25mm}+\frac{[f_{E'}-f_{E}]}{\omega^{2}_{\mu}+(E'-E)^{2}}
	N_{f}{\Big )}\;.
\label{cleanf}
\end{eqnarray}
Here $f_{E}\equiv f(E)$ is the Fermi function, and $S_{f}$ and $N_{f}$ are
\begin{eqnarray}\nonumber
	S_{f}=(E'+E)\Delta^{2}\;;\;\;N_{f}=(E'-E)\Delta^{2}\;.
\end{eqnarray}
For the other bubbles we obtain similar expressions with
\begin{eqnarray}\nonumber
	S_{g}&=&[(E'+E)(\xi\xi'-EE')+i\omega_{\mu}(\xi'E-\xi E')]\;,\\\nonumber
	N_{g}&=&[(E'-E)(\xi\xi'+EE')-i\omega_{\mu}(\xi'E+\xi E')]\;,\\\nonumber
	S_{k}&=&[(E'+E)\xi\Delta+i\omega_{\mu}E\Delta]\;,\\\nonumber
	N_{k}&=&[(E'-E)\xi\Delta-i\omega_{\mu}E\Delta]\;,\\\nonumber
	S_{h}&=&[-(E'+E)(\xi\xi'+EE')+i\omega_{\mu}(\xi'E+\xi E')]\;,\\
	N_{h}&=&[-(E'-E)(\xi\xi'-EE')-i\omega_{\mu}(\xi'E-\xi E')]\;.
\label{cleang}
\end{eqnarray}
The remaining integral over momenta can in general not be given in closed
form. Let us therefore consider the simple limits. For external momentum and
frequency much smaller than the energy gap and for $\omega_{\mu}\ll v_{F}q$,
the kernel $f_{0}$ reduces to
\begin{eqnarray}\nonumber
	f_{0}&=&N_{0}\int d\xi \frac{\Delta^{2}_{0}}{2E^{2}}
	\left(\frac{1}{2E}[1-2f_{E}]+\partial_{E}f(E)\right)+\cdots\\
	&=&\frac{N_{0}}{2}\frac{n_{s}}{n}+\cdots\;.
\label{fclean}
\end{eqnarray}
For the others we obtain
\begin{eqnarray}\nonumber
	g_{0}&=&-N_{0}+\frac{N_{0}}{2}\frac{n_{s}}{n}+\cdots=
		\frac{N_{0}}{2}\frac{-n-n_{n}}{n}+\cdots\\\nonumber
	h_{0}&=&-\frac{1}{\lambda}+\frac{N_{0}}{2}\frac{n_{s}}{n}
		+N_{0}\Gamma i\omega_{\mu}\cdots\\
	k_{0}&=& N_{0}\Gamma\Delta_{0}+\frac{N_{0}}{4}\frac{n_{s}}{n}
	\frac{i\omega_{\mu}}{\Delta_{0}}+\cdots\;.
\label{gclean}
\end{eqnarray}
The first order expansion of the kernels in frequency and momentum at zero
temperature reads
\begin{eqnarray}\nonumber
	f_{0}&=&\frac{N_{0}}{2}-\frac{N_{0}}{12\Delta_{0}^{2}}
	\left(\omega^{2}_{\mu}+\frac{v^{2}_{F}}{3}{\bf q}^{2}\right)
	+\cdots\\\nonumber
	g_{0}&=&-\frac{N_{0}}{2}+\frac{N_{0}}{12\Delta_{0}^{2}}
	\left(\omega^{2}_{\mu}-\frac{v^{2}_{F}}{3}{\bf q}^{2}\right)
	+\cdots\\\nonumber
	h_{0}&=&-\frac{1}{\lambda}+\frac{N_{0}}{2}+N_{0}\Gamma i\omega_{\mu}
	+\frac{N_{0}}{6\Delta_{0}^{2}}
	\left(\omega^{2}_{\mu}+\frac{v^{2}_{F}}{3}{\bf q}^{2}\right)+\cdots\\
	k_{0}&=&N_{0}\Gamma\Delta_{0}+\frac{N_{0}}{4}\frac{i\omega_{\mu}}
	{\Delta_{0}}\left[1-\frac{1}{6\Delta^{2}_{0}}\omega_{\mu}^{2}
	+\cdots\right]\;.
\label{coldclean}
\end{eqnarray}
In the normal metal $T>T_{C}$ and $\Delta_{0}\equiv 0$ the expressions
simplify considerably. For $q\ll 2k_{F}$ we have
\begin{eqnarray}\label{cleannm}
	f_{0}&=&k_{0}=0\\\nonumber
	g_{0}&=& -N_{0}\left(1-\frac{i\omega_{\mu}}{2v_{F}q}
	\ln\left[\frac{i\omega_{\mu}+v_{F}q}{i\omega_{\mu}-v_{F}q}
	\right]\right)=\\\nonumber
	\omega\!\!\!\!&&\!\!\!\!\ll v_{F}q:\;\;\approx -N_{0}
	\left(1-\frac{\pi}{2}\vert\frac{\omega_{\mu}}{v_{F}q}\vert+
	\left(\frac{\omega_{\mu}}{v_{F}q}\right)^{2}+\cdots\right)\\
	\omega\!\!\!\!&&\!\!\!\!\gg v_{F}q:\;\;\approx -\frac{N_{0}}{3}
	\left(\frac{v_{F}q}{\omega_{\mu}}\right)^{2}\nonumber\\
	h_{0}&=& N_{0}\left[\frac{\pi}{8}\nonumber
		\frac{\vert\omega_{\mu}\vert+\frac{7\zeta(3)v^{2}_{F}}
		{6\pi^{3}T}{\bf q}^{2}}{T}+i\Gamma\omega_{\mu}-
		\ln\left(\frac{2e^{\gamma}\omega_{D}}{\pi T}\right)\right] \;.
\end{eqnarray}

The bubbles are related to one another by the exact Ward identities, and also
by approximate identities that are good in the limit of low external
momenta. As an example we discuss the relation between $g_{3}$, $g_{2}$, and
$g_{0}$.
\begin{eqnarray}\nonumber
	g_{3}&=&\left\{({\bf p}\times{\bf q})^{2}/2q^{4}\right\}_{GG}\approx\\
	&\approx&\frac{q^{a}q^{b}}{q^{4}}\frac{1}{3}\delta^{ab}p^{2}_{F}\nonumber
	\left\{1\right\}_{GG}=\frac{p^{2}_{F}}{3{\bf q}^{2}}g_{0}\; ;\\\nonumber
	g_{2}&=&\left\{[ {\bf q}\cdot({\bf p}+{\bf q}/2) ]^{2}/
	q^{4}\right\}_{GG}\approx\\
	&\approx&\frac{q^{a}q^{b}}{q^{4}}\frac{1}{3}\delta^{ab}p^{2}_{F}
	\left\{1\right\}_{GG}=\frac{p^{2}_{F}}{3{\bf q}^{2}}g_{0}\;.
\label{connect}
\end{eqnarray}
Here we have used $(1/2)\int^{\pi}_{0}d\theta\sin\theta\cos^{2}\theta=1/3$ and
$(1/2)\int^{\pi}_{0}d\theta\sin\theta\sin^{2}\theta=2/3$ as well as the fact
that internal momenta can be taken at the Fermi energy. In doing so, one makes
a slight error and it can be shown that the leading term in the difference
$g_{3}-g_{2}$ is $N_{0}/12$, which is responsible for Landau diamagnetism as
discussed in Section III.

\subsection{Dirty Limit}
We will use the notation $\omega=\omega_{\nu}$, $\omega'=
\omega_{\nu}+\omega_{\mu}$, $W=\sqrt{\omega^{2}+\Delta^{2}_{0}}$, and
$W'=\sqrt{\omega'^{2}+\Delta^{2}_{0}}$. In the presence of impurities we
replace all frequencies and the gap by $\tilde{\omega}=\eta\omega$ and
$\tilde{\Delta}_{0}=\eta\Delta_{0}$, with $\eta=[1+1/(2\tau_{r}W)]$,
in the single particle Green's functions. In particular
$\tilde{W}=W+1/2\tau_{r}$\cite{agd}.

Furthermore, we will restrict ourselves to the limit $q\ll 2k_{F}$ and we put
$Q\equiv xv_{F}q$, with $x=\cos\theta$.
Although formally one should first sum over the internal frequency, in the
dirty limit it is more
convenient to integrate over energy first. By reversing the order one only
misses a constant $-N_{0}$ in the expression for $g_{0}$, which is added
by hand later on. The integral over energy $\xi_{p}$ by contour integration
is straightforward and the angular integration gives an $\arctan$. We find
for $f_{0}$
\begin{eqnarray}\label{dirt}
	f_{0}&=&\!\!\int\!{\,\mathchar'26\mkern-9mu{\!d}} p
	\frac{\tilde{\Delta}_{0}}{\tilde{\omega}^{2}_{\nu}+\xi^{2}_{\bf p}+
		\tilde{\Delta}^{2}_{0}}
	\frac{\tilde{\Delta}_{0}}{(\tilde{\omega}_{\nu}\!+
		\tilde{\omega}_{\mu})^{2}\!+
	\xi^{2}_{{\bf p}+{\bf q}}\!+\tilde{\Delta}^{2}_{0}}\\\nonumber
	&=&\pi N_{0}\frac{1}{\beta}\sum_{\omega_{\nu}}
	\frac{\Delta^{2}_{0}}{v_{F}qWW'}
	\arctan\left(\frac{v_{F}q}{W+W'+\tau^{-1}_{r}}\right).
\end{eqnarray}
In the dirty limit when $\Delta_{0}\tau_{r}\ll 1$, the $\arctan$ function can
now be expanded in $v_{F}q\tau_{r}=ql$ and $\omega_{\mu}\tau_{r}$, and to
leading order
\begin{eqnarray}\nonumber
	\frac{1}{v_{F}q}\arctan\left(\frac{v_{F}q}{W+W'+\tau^{-1}_{r}}\right)
	\approx\\\approx\frac{1}{W+W'+D{\bf q}^{2}+\tau^{-1}_{r}}\;,
\label{arctan}
\end{eqnarray}
where $D=\tau_{r}v^{2}_{F}/3$ is the diffusion constant. The full bubble
including the vertex correction due to impurity ladders is obtained from
self-consistency~\cite{fluct} by simply dropping the $\tau^{-1}_{r}$ from the
right hand side of Eq.~(\ref{arctan}). We find the final expression for the
disorder averaged polarization bubble
\begin{eqnarray}\nonumber
	f_{0}&=&\pi N_{0}\frac{1}{\beta}\sum_{\omega_{\nu}}
	\frac{\Delta^{2}_{0}}{WW'(W+W'+D{\bf q}^{2})}\;,
\label{dirtyf}
\end{eqnarray}
and analogously in the same limit $D{\bf q}^{2},\omega_{\mu}\Delta_{0}\lesssim
\tau^{-1}_{r}$ we obtain for the other kernels

\begin{eqnarray}\nonumber
	g_{0}&=&-N_{0}+\pi N_{0}\frac{1}{\beta}\sum_{\omega_{\nu}}\nonumber
	\frac{WW'-\omega\omega'}{WW'(W+W'+D{\bf q}^{2})}\;,\\ \nonumber
	h_{0}&=&\pi\frac{1}{\beta}\sum_{\omega_{\nu}}\nonumber
	\frac{-N_{0}(WW'+\omega\omega')+N'_{0}WW'i\tilde{\omega}_{\mu}}
	{WW'(W+W'+D{\bf q}^{2})}\;,\\
	k_{0}&=&\pi\frac{1}{\beta}\sum_{\omega_{\nu}}
	\frac{N_{0}(-i\omega\Delta_{0})+N'_{0}WW'\tilde{\Delta}}
	{WW'(W+W'+D{\bf q}^{2})}\;.
\label{dirtyghk}
\end{eqnarray}
A considerable simplification occurs for $T>T_{C}$ when $\Delta_{0}\equiv 0$.
In this case the remaining sums over the internal frequencies can be carried
out and yield differences of digamma functions. The low momentum and frequency
expansion gives
\begin{eqnarray}
	f_{0}&=&k_{0}=0\;,\label{dirtynm} \\\nonumber
	g_{0}&=& -N_{0}\frac{D{\bf q}^{2}}
		{\vert\omega_{\mu}\vert+D{\bf q}^{2}}\;,\\
	h_{0}&=& N_{0}\left[\frac{\pi}{8}\nonumber
		\frac{\vert\omega_{\mu}\vert+D{\bf q}^{2}}{T}+i\Gamma\omega_{\mu}
		-\ln\left(\frac{2e^{\gamma}\omega_{D}}{\pi T}\right)\right]\;.
\end{eqnarray}
In this limit, the kernels $g_{0}$ and $h_{0}$ are nothing but the Diffuson and
Cooperon. For temperatures $T<T_{C}$ no simple expressions are available.
However, the bubbles at zero external momentum and frequency are known
\begin{eqnarray}
	f_{0}(0)&=&\frac{\pi}{2}N_{0}\frac{1}{\beta}\sum_{\omega_{\nu}}
		\frac{\Delta^{2}_{0}}{(\omega^{2}_{\nu}+
		\Delta^{2}_{0})^{3/2}}=\frac{N_{0}}{2}\frac{n_{s}}{n}
\label{dirty2}
\end{eqnarray}
See Eq.~(\ref{gclean}) for the other kernels.

\subsection{Higher Order Bubbles}
In Section 3 we need the diagrams with 3 and 4 Green's functions at zero
external momenta for the construction of the Ginzburg-Landau functional.
The third order contribution $I^{-}_{3}$ is independent of the mean free path
\begin{eqnarray}\nonumber
	I^{-}_{3}&=&\int{\,\mathchar'26\mkern-9mu{\!d}}
	p[G(p)\bar{G}(p)G(p)-\bar{G}(p)G(p)\bar{G}(p)]\\
	&=&\frac{1}{\beta}\sum_{\omega_{\nu}}\int d\xi(N_{0}+\xi N'_{0}
	  +\cdots)\frac{-2\xi}{(\omega^{2}_{\nu}+\xi^{2})^{2}}\nonumber\\
	&=&-2N'_{0}\frac{1}{\beta}\sum_{\omega_{\nu}}\frac{1}{\vert
	  \omega_{\nu}\vert}\int^{\infty}_{-\infty}\frac{dxx^{2}}
	  {(1+x^{2})^{2}}\nonumber\\
	&=&-N'_{0}\ln\left(\frac{2e^{\gamma}\omega_{D}}{\pi T}\right)
	  \approx -N_{0}\Gamma\;,
\label{3bubb-}
\end{eqnarray}
and is proportional to the particle-hole asymmetry of the problem, whereas the
combination $I^{+}_{3}$ is
\begin{eqnarray}\nonumber
	I^{+}_{3}&=&\int{\,\mathchar'26\mkern-9mu{\!d}}
	p[G(p)\bar{G}(p)G(p)+\bar{G}(p)G(p)\bar{G}(p)]\\
	&=&\frac{1}{\beta}\sum_{\omega_{\nu}}\int d\xi N_{0}(\xi)
	\frac{2i\omega_{\nu}}{(\omega^{2}_{\nu}+\xi^{2})^{2}}=0\;.
\label{3bubb+}
\end{eqnarray}
The fourth order contribution is 
\begin{eqnarray}\nonumber
	I_{4}&=&\int{\,\mathchar'26\mkern-9mu{\!d}}
	p G^{2}(p)\bar{G}^{2}(p)=N_{0}\frac{1}{\beta}
	\sum_{\omega_{\nu}}\int d\xi\frac{1}{(\omega^{2}_{\nu}+\xi^{2})^{2}}\\
	&=&N_{0}\frac{1}{\beta}\sum_{\omega_{\nu}}\frac{1}
	{\vert\omega_{\nu}\vert^{3}}\int^{\infty}_{-\infty} d\xi
	\frac{dx}{(1+x^{2})^{2}}\nonumber\\
	&=&N_{0} \frac{7\zeta(3)}{4\pi^{3}T^{2}}\frac{\pi}{2}\equiv N_{0}b
	\;;\;\; b=\frac{7\zeta(3)}{8\pi^{2}T^{2}}\;.
\label{4bubb}
\end{eqnarray}
Here we have introduced the usual Ginzburg-Landau parameter $b$.
Finally, below $T_{C}$ we need the combination
\begin{eqnarray}\nonumber
	I^{\rm sc}_{3}&=&\!\!\int\!\!{\,\mathchar'26\mkern-9mu{\!d}}
	p[G(p)-\bar{G}(p)][3F(p)F(p)+G(p)\bar{G}]
	\\ &\approx& -2N_{0}\Gamma \;.
\label{3bubb_sc}
\end{eqnarray}

\end{document}